\pdfoutput=1
\documentclass[conference]{IEEEtran}
\usepackage{amsmath,amssymb,amsfonts,mathrsfs,bm}
\usepackage{amstext}
\usepackage{upgreek}
\usepackage{multicol}
\usepackage{indentfirst}
\usepackage{graphicx}
\usepackage{paralist}
\usepackage{multirow}
\usepackage{tabularx}
\usepackage[noadjust]{cite}

\usepackage{booktabs}
\usepackage{subfigure}
\usepackage{color}

\IEEEoverridecommandlockouts
\allowdisplaybreaks
\usepackage{algorithm,algpseudocode,float}
\floatname{algorithm}{\small Algorithm}
\makeatletter
\newcommand{\algmargin}{\the\ALG@thistlm}
\makeatother
\algnewcommand{\parState}[1]{\State%
	\parbox[t]{\dimexpr\linewidth-\algmargin}{\strut #1\strut}}

\newtheorem{proposition}{\bf Proposition}

\usepackage{accents}

\allowdisplaybreaks
\begin{document}
\title{A Learning-based Approach to Joint Content Caching and Recommendation at Base Stations \vspace{-3.5mm}}
\author{
 	\IEEEauthorblockN{{Dong Liu and Chenyang Yang}}\\
 		\vspace{-4mm}
	\IEEEauthorblockA{Beihang University, Beijing, China\\
	Email: \{dliu, cyyang\}@buaa.edu.cn}
	\thanks{This work was supported in part by the  National Natural Science Foundation of China (NSFC) under Grant	61731002 and  61429101.}
		\vspace{-8.5mm}
}
\maketitle

\begin{abstract}
Recommendation system is able to shape user demands, which can be used for boosting caching gain. In this paper, we jointly optimize content caching and recommendation at base stations to maximize the caching
gain meanwhile not compromising the user preference. We first propose a model to capture the impact of recommendation on user demands, which is  controlled by a user-specific psychological threshold. We then formulate a joint caching and recommendation problem maximizing the successful offloading probability, which is a mixed integer programming problem. We develop a hierarchical iterative algorithm to solve the problem when the threshold is known. Since the user threshold is unknown in practice, we proceed to propose an $\varepsilon$-greedy algorithm to find the solution by learning the threshold via interactions with users. Simulation results show that the proposed algorithms improve the successful offloading probability compared with prior works with/without recommendation. The $\varepsilon$-greedy algorithm learns the user threshold quickly, and achieves more than $1-\varepsilon$ of the performance obtained by the algorithm with known threshold.
\end{abstract}

\section{Introduction}
%Motivated by the fact that a large portion of mobile data traffic is generated by many duplicate downloads of a few popular contents [5],
Caching at the base stations (BSs) has been acknowledged as a promising way to support the explosively increasing traffic demands and improve user experience \cite{liu2016caching}.
To increase the caching gain at the wireless edge where each node is with limited storage-size,
various proactive caching solutions have been proposed \cite{golrezaei2012femtocaching,Blaszczyszyn2015optimal,TMX}.

%Based on multi-armed bandit theory, caching policy was optimized \cite{MAB} after learning content popularity.

%Using collaborative filtering tools, user preference is exploited in \cite{} to increase the offloading ratio of cache-enabled device-to-device network.

A ``more proactive'' approach to increase caching gain is to shape user demand itself to be more caching-friendly. Recommender system, whose original goal is to relieve users from information overload by recommending the preferred contents to each user  \cite{ekstrand2011collaborative}, has been leveraged for demand-shaping in content distribution networks (CDN) \cite{Krishnappa,Shaping} as well as wireless networks \cite{guo,nudge,soft}.

The basic idea of demand-shaping is that the recommender system does not necessarily recommend the contents that best match the taste of each individual user; instead, it can recommend the contents that match the user preference adequately and are also attractive to other users. If the users would accept such recommendation, the user demands will be less heterogeneous, which makes the content popularity more skewed.  It has been observed in \cite{Krishnappa} that the hit ratio of YouTube's caches was increased by reordering the related video list so that the already cached contents are at the top of the list. In \cite{Shaping}, the cost of service provider for pushing contents to users was reduced, where demand-shaping was achieved by adjusting the rating of contents shown to each user.

In prevalent mobile networks, recommendation and caching are managed by different entities, i.e., content providers and mobile network operators (MNOs), respectively. Yet content caching and recommendation are coupled with each other, since recommendation influences user demands, which further affects caching policy. Therefore, optimizing caching or recommendation alone cannot fully reap their benefits. Recently, there is a trend towards the convergence of MNOs and content providers. In \cite{guo},
recommendation was integrated with wireless edge caching, where the BS recommends all cached contents to every user so that the content popularity is more skewed and hence achieves higher caching gain. However, the preference of each user was assumed as identical to the content popularity. Considering heterogeneous user preference, a personalized recommendation policy was proposed in  \cite{nudge} to improve cache-hit ratio by recommending contents that are both cached and appealing to each user. In particular, the BS first optimizes caching
policy based on recommending top-$N$ contents according
to the preference of each user, and then adjusts
the recommendation lists based on the cached contents. In \cite{soft}, caching policy was optimized to maximize a ``soft" cache-hit ratio by offering related contents in the cache if the originally requested content is not cached in nearby BSs. 
However, the user demands after recommendation are assumed known in \cite{nudge} and \cite{soft} (i.e., the probability of user requesting for each content after recommendation \cite{nudge}, or the probability of user accepting a related content recommendation \cite{soft} are known), which are unavailable in practice. Besides, existing works of integrating recommendation into wireless networks \cite{guo,nudge,soft} do not take fading channels and interference into account, which also affect caching gain and user experience.

In this paper, we jointly optimize content caching and recommendation at BSs. To capture the impact of recommendation on user preference, we propose a model for user demands with recommendation, which is controlled by a user-specific threshold determining whether the user is prone to request the recommended contents. We then formulate a joint caching and recommendation problem maximizing the successful offloading probability, defined as the probability that a request can be offloaded to the cache with received signal-to-interference ratio (SIR) larger than a given value. To compare fairly with prior works, we first propose a hierarchical iterative algorithm to solve the problem with known user threshold. Then, we propose an $\varepsilon$-greedy algorithm to find the solution with unknown threshold, inspired by the concepts in reinforcement learning \cite{sutton1998reinforcement}. The algorithm can trade off between learning the user threshold via interactions with users (i.e., exploration) and maximizing the performance based on the currently estimated threshold (i.e., exploitation).

The rest of the paper is organized as follows. In section II, we present the system model. In section III, we introduce the user demands model before and after recommendation. In section IV, we formulate the joint optimization problem, and propose algorithms to find the solutions with known and unknown user threshold. Simulation results are provided in section IV, and section V concludes the paper.

\section{System Model}
Consider a large-scale cache-enabled cellular network, where each BS is equipped with a cache and connected to a central unit (CU) in the core network via backhaul link.

Suppose that there are $N_u$ users in the network, which may request contents from a catalog containing $N_f$ equal-sized contents. Each BS is equipped with $N_t$ antennas, and each user is with a single antenna. The spatial distribution of BSs are modeled as homogeneous Poisson point process (PPP) with density $\lambda$ denoted as $\Phi$. Since proactive
caching decisions are made during off-peak time, which might be hours in advance of the time
of delivering content, the location where a mobile user will initiate a request is hard to predict in advance. To reflect the impact of location uncertainty on optimizing the policy, we assume that the users are uniformly located in the network.

Each BS can cache at most $N_c$ contents, and recommend $N_m$ contents to each user, where $N_m \ll N_c $ since a user usually would not like to read a long list of recommendation. We consider probabilistic caching policy, where each BS
independently selects contents to cache according to a probability distribution \cite{Blaszczyszyn2015optimal}. To shape the user demands without comprising user preference, we consider personalized recommendation.  In each time slot (e.g., few hours or a day), the CU optimizes the caching policy and recommendation policy, and broadcasts the obtained policies to each BS.

Denote the caching policy in the $t$th time slot as $\mathbf c^{(t)} = [c_{1}^{(t)}, \cdots, c_{N_f}^{(t)}]^T$, where $c_{f}^{(t)}\in [0, 1]$ is the probability that a BS caches the $f$th content. With given $\mathbf c^{(t)}$, each BS can determine which contents to cache by the method in \cite{Blaszczyszyn2015optimal}. The distribution of the BSs caching the $f$th content can
be regarded as a thinning of $\Phi$ with probability $c_{f}$, which also follows a homogeneous PPP with density $c_f\lambda$ (denoted by $\Phi_f$). The distribution of the BSs not caching the $f$th content follows homogeneous PPP with density
$(1-c_f)\lambda$ (denoted as $\Phi'_f$).

Denote the recommendation policy in the $t$th time slot as $\mathbf M^{(t)} = [\mathbf{m}_1^{(t)}, \cdots, \mathbf{m}_{N_u}^{(t)}]^T$, where $\mathbf{m}_{u}^{(t)} = [m_{u1}^{(t)},\cdots,m_{uN_f}^{(t)}]^T$ is the policy to the $u$th user, and $\sum_{f = 1}^{N_f}m_{uf}^{(t)} = N_m$. If the $f$th content is recommended to the $u$th user, $m_{uf}^{(t)}= 1$, otherwise, $m_{uf}^{(t)} = 0$. The recommended contents to the $u$th user, i.e., recommendation list, is denoted as a set $\mathcal{M}_u \triangleq \{f|m_{uf}^{(t)} = 1\}$.

When a user intends to request a content (e.g., open a video application (app) on mobile device), the recommendation list is first presented (e.g., shown on the home screen of the app). If the user has no strong preference towards a specific content and the recommended contents match the taste of user adequately, the user is more likely to click a content in the recommendation list to initiate a request. By contrast, if the user has already determined what content to watch or the recommended contents do not match the taste of the user at all, the user may simply ignore the recommendation.

Suppose that the eventually requested content is the $f$th content. If the content is cached at one or more BSs nearby the user and the corresponding received SIR\footnote{We consider an interference-limited network, where BSs are densely deployed and hence the noise power can be neglected.} exceeds a value $\gamma_0$, the content will be downloaded to the user from the nearest one of these BSs. Then, the request can be successfully offloaded to the cache. Otherwise, the content is downloaded to the user via local BS  (i.e., the nearest BS to the user) from the backhaul.
Assume that each BS serves every $N_t$ users in the same time-frequency resource by zero-forcing beamforming with equal power allocation. Then, when the user requests the $f$th content and downloads from the closest BS $b_0$ that caches the $f$ content, the SIR can be expressed as
\begin{equation}\label{SIR}
\gamma_f \!=\! \frac{\frac{P}{N_t}hr^{-\alpha}}{\sum_{b\in\Phi_f\! \backslash b_0}P g_b r_b^{-\alpha} \!+\! \sum_{b\in\Phi'_{f}} Pg_b r_b^{-\alpha}} \! \triangleq \!\frac{hr^{-\alpha}}{N_t(I_f \! + \! I'_{f})} \!\!
\end{equation}
where $P$ is the transmit power of BS, $h$ is the equivalent channel power (including channel
coefficient and beamforming) from the associated BS $b_0$ to the user, $r$ is the corresponding distance, $\alpha$ is the pathloss exponent, $I_f \triangleq \sum_{b\in\Phi_f\! \backslash b_0}g_b r_b^{-\alpha}$ is the normalized interference from the BSs that cache the $f$th content and $I_f' \triangleq \sum_{b\in\Phi'_{f}} g_b r_b^{-\alpha} $ is the normalized interference
from the BSs that do not cache the $f$th content, $g_b$ and $r_b$ are respectively the equivalent interference channel power and distance from the $b$th BS to the user. We consider Rayleigh fading channels. Then, $h$ follows exponential distribution with unit mean, i.e., $h \sim \exp(1)$, and $g_b$ follows gamma distribution with shape parameter $N_t$ and unit mean, i.e., $g_b \sim \mathbb{G}(N_t , 1/N_t )$ \cite{adhoc}.

\section{Modeling User Demands Before and After Recommendation}
In this section, we first introduce inherent user demands before a user sees the recommendation list. Then, we provide a model to reflect user personality in terms of how easy a user is influenced by recommendation.
\subsection{Inherent User Preference}
Each content can be represented by a $K$ dimensional feature vector, which can be extracted from the content metadata in the form of tags (e.g, the genre of a movie or a song),  or learned with various representation learning methods such as matrix factorization \cite{ekstrand2011collaborative} and deep neural networks \cite{deep}. Denote the feature vector of the $f$th content as $\mathbf{x}_f = [x_{f1}, \cdots, x_{fK}]^T$, where $x_{fk}$ reflects the relevance of the $f$th content to the $k$th feature. Similarly, the $u$th user can be represented by a $K$ dimensional feature vector $\mathbf{y}_u = [y_{u1}, \cdots, y_{uK}]^T$, where $y_{uk}$ reflects the interest of the $u$th user to the $k$th feature. Since the user's interest to each feature changes slowly with respect to the  time slot duration, the user feature vector can be learned from its content request history. % or adopted from other domain by transfer learning approach \cite{}.
Then, the inner product $\mathbf x_{f}^T\mathbf y_{u} $ reflects the attractiveness of the $f$th content to the $u$th user.

The \emph{inherent user preference} is the probability distribution of user requests for every content without the influence of recommendation. Denote $p_{uf}$ as the probability that the $u$th user requests the $f$th content conditioned on that the user requests a content before recommendation. Based on the multinomial logit model in discrete choice theory \cite{greene2003econometric}, which is often used in economics to describe, explain, and predict choices among multiple discrete alternatives, the user preference before recommendation can be obtained as
$p_{uf} = \frac{\exp(\mathbf x_{f}^T\mathbf y_{u}) }{\sum_{f'=1}^{N_f}\exp(\mathbf x_{f'}^T\mathbf y_{u}) }$.
Such a function is also acknowledged as a softmax function in machine learning field.

\subsection{User Preference After Recommendation}\label{III.B}
To capture the impact of recommendation on the preferences of different users, we introduce a psychological threshold  $\theta_u$ ($0\leq \theta_u \leq 1$) to reflect the personality of the $u$th user in terms of the likelihood to accept a recommendation. Specifically, when the inherent preference is above the threshold, i.e., $p_{uf} \geq \theta_u$, which means that the $f$th content attracts the $u$th user sufficiently, the user will regard the $f$th content as a \emph{candidate content} to request if the content is recommended to the user. We call the contents in the recommendation list $\mathcal{M}_u$ that satisfies $p_{uf} \geq \theta_u$ as the \emph{candidate subset}, denoted as $\mathcal{A}_{u} \triangleq \{f ~|~ p_{uf} \geq \theta_u, f \in \mathcal{M}_u\}$. The candidate subset $\mathcal{A}_{u}$ restricts the contents that the $u$th user may request from $\mathcal{M}_u$.

Denote the probability that the $u$th user is influenced
by the recommendation as $q_u$. Intuitively, a recommendation list will be appealing to a user if the list includes many contents sufficiently attractive to the user (i.e., $\mathcal{A}_u$ is large). On the other hand, if the list includes too many contents (i.e., $N_m$ is large),  the user may ignore the recommendation due to information overload. Therefore, it is reasonable to assume that $q_u = \frac{|\mathcal{A}_u|}{N_m}$, which can be obtained after $\theta_u$ is known, where $|\cdot|$ denotes the cardinality of a set. If the request of the $u$th user is influenced by recommendation, the user will request content (say the $f$th content) from the candidate subset $\mathcal{A}_u$ with probability
%\begin{equation}
$\tilde p_{uf} = \left\{\begin{array}{ll}
 \frac{\exp(\mathbf{x}_f^T \mathbf{y}_u )}{\sum_{f'\in \mathcal{A}_u} \exp(\mathbf{x}_{f'}^T \mathbf{y}_u)}, ~f\in \mathcal{A}_u \\
 0, ~f\notin \mathcal{A}_u
\end{array}
\right.$.
%\label{eqn:puf1}
%\end{equation}

\begin{figure}[!htb]
	\centering
	\includegraphics[width=0.38\textwidth]{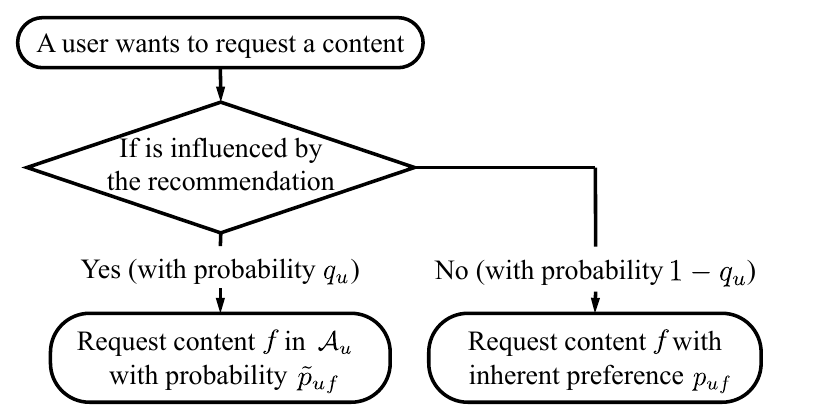}
	\vspace{-1mm}
	\caption{A generative model of user demands with recommendation.} \label{fig:rec}
	\vspace{-2mm}
\end{figure}

The $u$th user is not influenced by recommendation with probability $1-q_u$. For such a user, it requests a content according to its inherent preference. The generative process of the user demands after recommendation is shown in Fig.~\ref{fig:rec}.

Then, we can obtain the \emph{user preference after recommendation} based on the law of total probability. Specifically, the probability that the $u$th user requests the $f$th content after the recommendation can be expressed as
\begin{align}
q_{uf}(\mathbf{m}_u^{(t)}, \theta_u)   = &~ q_u \tilde{p}_{uf} + (1-q_u) p_{uf} \nonumber \\
  = & ~\tfrac{\sum_{f'=1}^{N_f} a_{uf'}m_{uf'}^{(t)}}{N_m} \tfrac{a_{uf} m_{uf}^{(t)}  p_{uf}}{\sum_{f' = 1}^{N_f} a_{uf'} m_{uf'}^{(t)}p_{uf'}} \nonumber \\
&~+ \Big(1- \tfrac{\sum_{f'=1}^{N_f} a_{uf'}m_{uf'}^{(t)}}{N_m}\Big) p_{uf} \label{eqn:quf}
\end{align}
where $a_{uf} \!=\! 1$ if $p_{uf} \geq \theta_u$, and $a_{uf} \!=\!0$ otherwise.

From the above model, we can see that if the user threshold is too high or the recommended contents do not match the inherent user preference adequately, the user's  request will not be affected by recommendation (i.e., $\mathcal{A}_{u}$ will be empty and hence $q_u = 0$). On the contrary, if the user threshold is low  or all the recommended contents match the taste of the user sufficiently so that $|\mathcal{A}_u| = N_m$ and hence $q_u = 1$, the user will only request content in the recommendation list. Considering that $N_m \ll N_f$ in reality, the number of possible contents that a user may request shrink significantly compared with the case without recommendation.

Different from the model introduced in \cite{nudge} that recommendation will always boost the request probability of each user for every recommended content equally, our model captures the fact that the effectiveness of recommendation depends on both the attractiveness of the recommended contents to the user (i.e., $p_{uf}$), and the personality of the user (reflected by the psychological threshold $\theta_u$). In practice, $\theta_u$ is never known, which can be learned via interactions with the user. Then, $\mathcal{A}_u$, $q_u$, and $\tilde p_{uf}$ can be obtained accordingly, with which the user preference after recommendation can be updated by \eqref{eqn:quf}. 

To demonstrate the gain from joint optimization of caching and recommendation by learning the user threshold, we assume that $\mathbf x_f$ and $\mathbf y_u$ are learned  perfectly and hence the inherent user preference $p_{uf}$ is known \emph{a priori}. We will investigate the joint learning user preferences before and after recommendation in future work.

\section{Joint Content Caching and Recommendation}
In this section, we formulate a joint content caching and recommendation problem, and solve the problem with known and unknown user threshold, respectively.

\subsection{Problem Formulation}
To reflect the offload gain, we consider \emph{successful offloading probability}, defined as the probability that the requested content can be downloaded from the cache with received SIR larger than $\gamma_0$. Based on the law of total probability, the successful offloading probability after recommendation is
\begin{equation}
\mathsf{s}(\mathbf{c}^{(t)}, \mathbf{M}^{(t)}, {\boldsymbol{\theta}}) = \sum_{u=1}^{N_u}v_u \sum_{f=1}^{N_f} q_{uf}(\mathbf{M}^{(t)}, {\boldsymbol{\theta}}) \mathbb{P}(\gamma_f \geq \gamma_0) \label{eqn:s}
\end{equation}
where ${\boldsymbol{\theta}} = [\theta_1, \cdots, \theta_{N_u}]$, $v_u$ is the probability that the request is sent from the $u$th
user, which reflects the activity level of the user,
$\mathbb{P}(\gamma_f \geq \gamma_0) =  \frac{c_f^{(t)}}{\mathsf{G}_1(\gamma_0)c_f^{(t)} + \mathsf{G}_2(\gamma_0)}$
is derived in Appendix~A with $\mathsf{G}_1(\gamma_0) = {}_{2}F_1[-\frac{2}{\alpha}, N_t; 1-\frac{2}{\alpha}, -\gamma_0 ] - \Gamma (1-\frac{2}{\alpha}) \Gamma(N_t + \frac{2}{\alpha}) \Gamma(N_t)^{-1}\gamma_0^{\frac{2}{\alpha}}$ and $\mathsf{G}_2(\gamma_0) = \Gamma (1-\frac{2}{\alpha}) \Gamma(N_t + \frac{2}{\alpha}) \Gamma(N_t)^{-1}\gamma_0^{\frac{2}{\alpha}}$.

%We can rewritten \eqref{eqn:s} into $\mathsf{s}(\mathbf{c}^{(t)}, \mathbf{M}^{(t)}, {\boldsymbol{\theta}}) = \sum_{f=1}^{N_f} (\sum_{u=1}^{N_u}v_u q_{uf}(\mathbf{M}^{(t)}, {\boldsymbol{\theta}})) \mathbb{P}(\gamma_f \geq \gamma_0)$, where the term $\sum_{u=1}^{N_u}v_u q_{uf}(\mathbf{M}^{(t)}, {\boldsymbol{\theta}})$ is actually the content popularity after recommendation, i.e., the probability that a requested content from the $N_u$ users is the $f$th content after recommendation. This suggests that the successful offloading probability depends on the aggregated user demands rather than the individual user preference.

Then, the joint content caching and recommendation problem can be formulated as
\begin{subequations}
	\begin{align}
	 \mathsf{P}_1: ~\max_{\mathbf c^{(t)}, \mathbf{M}^{(t)}} ~& {\sf s}(\mathbf c^{(t)}, \mathbf{M}^{(t)},{\boldsymbol{\theta}}) \\
	s.t. ~& \sum_{f = 1}^{N_f} c_f^{(t)} \leq N_c \label{eqn:con1}\\
	& \sum_{f = 1}^{N_f} m_{uf}^{(t)} = N_m,~\forall u \label{eqn:con2}\\
	& 0\leq c_f^{(t)} \leq 1,~ m_{uf} \in \{0, 1\},~\forall u, f \label{eqn:con3}
	\end{align}
\end{subequations}
where \eqref{eqn:con1} is equivalent to the cache size constraint
\cite{Blaszczyszyn2015optimal}, \eqref{eqn:con2} is the recommendation list constraint, and \eqref{eqn:con3} is the cache probability and recommendation decision variable constraint.

Problem $\mathsf{P}_1$ is a mixed integer programming problem involving binary variables $m_{uf}^{(t)}$ and continuous variable $c_{f}^{(t)}$.
When ${\boldsymbol{\theta}}$ is known, we propose a hierarchical iterative algorithm, where the policies can be obtained for every time slot independently. In the outer iteration, a greedy algorithm is used to find the recommendation policy. In the inner iteration, the optimal caching policy can be found by bisection searching of a scalar with a given recommendation policy.  When ${\boldsymbol{\theta}}$  is unknown, we propose an $\varepsilon$-greedy algorithm, where the threshold is
estimated by observing the reaction of each user in each time slot after providing a recommendation and then updating the recommendation.

\subsection{Hierarchical Iterative Algorithm for Known ${\boldsymbol{\theta}}$}
\subsubsection{Inner Iteration}
With any given recommendation policy $\mathbf{M}^{(t)}$, similar to the derivation in \cite{liu2017caching}, it is not hard to prove that $\mathsf{P}_1$ is concave in $\mathbf c^{(t)}$. Then, the optimal caching policy can be obtained from the Karush-Kuhn-Tucker condition as
\begin{equation}
\tilde c_{f}^{(t)}(\mathbf{M}^{(t)}, {\boldsymbol{\theta}})\! =\! \bigg[\tfrac{\sqrt{\mathsf{G}_2(\gamma_0)\sum_{u=1}^{N_u}v_uq_{uf}(\mathbf{m}_u^{(t)}, \theta_u)}}{\sqrt{\mu} \mathsf{G}_1(\gamma_0)} - \tfrac{\mathsf{G}_2(\gamma_0)}{\mathsf{G}_1(\gamma_0)}\bigg]_0^1 \!\! \label{eqn:c}
\end{equation}
where $[x]_0^1  = \max\{\min\{x, 1\}, 0\}$ denotes that $x$ is truncated by $0$ and $1$, and the Lagrange multiplier $\mu$ satisfying $\sum_{f=1}^{N_f} \tilde c_{f}^{(t)}(\mathbf{M}^{(t)}, {\boldsymbol{\theta}}) = N_c$ can be found by bisection searching. Considering that $0\leq \tilde c_{f}^{(t)}(\mathbf{M}^{(t)}, {\boldsymbol{\theta}})\leq 1$, we can obtain $0<\mu\leq \delta_0$ where $\delta_0 = \max_f\{\sum_{u=1}^{N_u}  v_u q_{uf}(\mathbf{m}_u^{(t)} , \theta_u)\} /\mathsf{G}_2(\gamma_0)$. Hence, the computation complexity of $\mu$ is $\mathcal{O}(\log_2(\delta_0/\delta))$ with error tolerance~$\delta$. It is noteworthy that $\sum_{u=1}^{N_u}v_uq_{uf}(\mathbf{m}_u^{(t)}, \theta_u)$ is actually the content popularity of the $f$th content after recommendation.

\subsubsection{Outer Iteration}
With the optimal caching policy in \eqref{eqn:c}, $\mathsf{P}_1$ degenerates into the optimization of $\mathbf M^{(t)}$ as $\max_{\mathbf{M}^{(t)}} {\sf s}(\tilde{\mathbf{c}}^{(t)}(\mathbf{M}^{(t)}, {\boldsymbol{\theta}}), \mathbf{M}^{(t)},{\boldsymbol{\theta}})$, where $\tilde{\mathbf{c}}^{(t)}(\mathbf{M}^{(t)}, {\boldsymbol{\theta}}) = [\tilde c_{1}^{(t)}(\mathbf{M}^{(t)}, {\boldsymbol{\theta}}), \cdots, \tilde c_{N_f}^{(t)}(\mathbf{M}^{(t)}, {\boldsymbol{\theta}})]$. The optimal  $\mathbf M^{(t)}$ can be found via exhaustive searching over $\binom{N_f}{N_m}{}^{N_u}$ possible candidates, which, however, is of prohibitive complexity.

In the following, we use greedy algorithm to find the recommendation policy. Since recommending the contents that a user is highly unlikely to request almost does not affect user demands, we can introduce a potential recommendation set for the $u$th user, $\mathcal{M}'_u$, as the $N_a$ contents with the largest values of $p_{uf}$, where $N_a \ll N_f$. In this way, we can control the deviation from the goal of
recommendation system (i.e., recommending the most preferred contents to the
user) such that the user can   indeed be satisfied with the recommendation.

In the algorithm, we first set the initial recommendation lists for all the users as empty at the start of each time slot. Then, at each iteration, we add one content to one user's recommendation list from its potential recommendation set so that the successful offloading probability is maximized. The iteration finally stops when each user is recommended with $N_m$ contents.
The whole procedure is shown in Algorithm \ref{alg:1}.

\begin{algorithm}
	\caption{\small Hierarchical iterative algorithm at time slot $t$}
	\label{alg:1}
	\renewcommand{\algorithmicrequire}{\small \textbf{Input:}}
	\renewcommand{\algorithmicensure}{\small \textbf{Output:}}
	\begin{algorithmic}[1] \small
		\Require $\{p_{uf}\}$, $\{v_u\}$, $\boldsymbol{\theta}$
		\State Initialize $\mathbf M^{(t)} = \mathbf{0}$,  potential recommendation set $\mathcal{M}'_u$ as the $N_a$ contents with the largest values of $p_{uf}$, and  $\mathcal{U} = \{1, \cdots, N_u\}$.
		\While{$\mathcal{U}$ is not empty}
		\parState{$[u^*, f^*] = \arg \max_{u \in \mathcal{U}, f \in \mathcal{M}'_u} {\sf s}(\tilde{\mathbf c}^{(t)}(\mathbf{M}^{(t)} + \Delta_{uf},{\boldsymbol{\theta}}), $ $\mathbf{M}^{(t)} + \Delta_{uf},{\boldsymbol{\theta}})$, and $\Delta_{uf}$ is a $N_u \times N_f$ dimensional $0$-$1$ matrix with only one ``$1$'' elements on the $u$th row and the $f$th column.}
		\State $\mathbf{M}^{(t)} \leftarrow \mathbf{M}^{(t)} + \Delta(u^*, f^*)$
		\State $\mathcal{M}'_u \leftarrow \mathcal{M}'_u \backslash f^*$\
		\If{$|\mathcal{M}_u| = N_m$}
	\State $\mathcal{U} \leftarrow \mathcal{U}\backslash u^*$
		\EndIf
		\EndWhile
		\Ensure The recommendation  policy, $\mathbf M^{(t)}$, and the caching policy $\mathbf{c}^{(t)} = \tilde{\mathbf{c}}^{(t)}(\mathbf M^{(t)}, {\boldsymbol{\theta}})$, for time slot $t$.
	\end{algorithmic}
\end{algorithm}

The search space for $u^*$ and $f^*$ in step (3) of Algorithm 1 is $|\mathcal{U}||\mathcal{M}'_u|$, and the complexity for computing $\tilde{c}^{(t)}(\mathbf{M}^{(t)} + \Delta_{uf},{\boldsymbol{\theta}})$ is $\mathcal{O}(\log_2(\delta_0/\delta))$ due to the bisection searching of $\mu$. Since $|\mathcal{U}||\mathcal{M}'_u| \leq N_u N_a$ and the algorithm stops with $N_u N_m$ iterations, the overall complexity of Algorithm 1 is at most $\mathcal{O}( N_u^2 N_m N_a\log_2(\delta_0/\delta))$, which is much lower than solving Problem 1 by exhaustive searching, i.e., $\mathcal{O}(\binom{N_f}{N_m}{}^{N_u}\log_2(\delta_0/\delta))$.

%We refer to this algorithm as the \emph{low-complexity} version of Algorithm 1.

We call the caching and recommendation policy obtained by Algorithm 1 with known user threshold as \emph{oracle policy}.

\subsection{$\varepsilon$-Greedy Algorithm for Unknown ${\boldsymbol{\theta}}$}
Without loss of generality, we rank the inherent preference of the $u$th user for the $N_f$ contents in ascending order as $p_{uf_1}\leq p_{uf_2}\leq \cdots \leq p_{uf_{N_f}}$, and divide $[0,1]$ into $N_f + 1$ intervals as $[0, p_{uf_{1}}], (p_{uf_{1}}, p_{uf_{2}}],\cdots, (p_{uf_{N_f}}, 1]$.

{\bf Remark 1:}
 The user demands after recommendation only depend on which interval $\theta_u$ falls in rather than the exact value of $\theta_u$. Define $\tilde\theta_u \triangleq p_{uf_{n}}\geq \theta_u > p_{uf_{n-1}}$\footnote{We define $p_{uf_{0}} \triangleq 0$ and $p_{uf_{N_f+1}} \triangleq 1$ to ensure mathematics rigorous.} as the  right end-point of the interval that $\theta_u$ lies in, and $\tilde{\boldsymbol{\theta}} \triangleq [\tilde \theta_1, \cdots, \tilde \theta_{N_u}]$. Then, we can obtain $q_{uf}(\mathbf{M}^{(t)}, \tilde\theta_u) = q_{uf}(\mathbf{M}^{(t)},  \theta_u)$ for $\forall~\mathbf{M}^{(t)}$,
i.e.,  the policy obtained by Algorithm 1 based on $ \tilde{\boldsymbol{\theta}}$ is the same as the oracle policy.

{\bf Remark 2:}
Define an indicative function,  $I(u,f) = 1$ if the $u$th user requests the $f$th content from the recommendation list,  $I(u,f) = 0$ otherwise. From the user demands model after recommendation, we can infer that $ \theta_u \leq p_{uf}$ if $I(u,f)= 1$.

Denote $\hat{\theta}_u^{(t)}$ as the estimate of $\theta_u$ in the $t$th time slot, and $\hat {\boldsymbol{\theta}}^{(t)} \triangleq [\hat \theta_1^{(t)},\cdots, \hat\theta_{N_u}^{(t)}]$.
To estimate ${\boldsymbol{\theta}}$, the initial estimate can be set as its upper bound, i.e.,  $\hat {\boldsymbol{\theta}}^{(1)} = \mathbf 1$. After observing a request of the $u$th user for the $f$th content in the $t$th time slot, its threshold can be updated as $\hat\theta_u^{(t+1)} = p_{uf}$ if $I(u,f) = 1$ and $\hat\theta_u^{(t)} > p_{uf}$ according to Remark 2. Then, at each time slot the CU computes the recommendation and caching policy using Algorithm 1 based on the current estimate $\hat {\boldsymbol{\theta}}^{(t)}$, and informs each BS of the recommendation and caching policy. This is actually \emph{exploitation} since the successful offloading probability is maximized based on the current knowledge of ${\boldsymbol{\theta}}$. However, if we always recommend contents with $\hat {\boldsymbol{\theta}}^{(t)}$, the recommendation will tend to be conservative (i.e., only recommend contents that best match the user's inherent preference) since $\hat \theta_u^{(t)} \geq \tilde \theta_u$ due to the initialization of $\hat{\boldsymbol{\theta}}^{(t)}$. Then, the estimated user threshold cannot converge to $\tilde {\boldsymbol{\theta}}$, which prevent the caching and recommendation policy to converge to the oracle policy eventually. Therefore, it is necessary to recommend contents that are not given by Algorithm 1, i.e., resorting to \emph{exploration} for improving the estimate of ${\boldsymbol{\theta}}$.

Exploitation is the right thing to do to maximize the performance in one time slot, but exploration may give better performance in the long run. Inspired by the trial-and-error approach to balance between exploration and exploitation in reinforcement learning \cite{sutton1998reinforcement}, we propose an $\varepsilon$-greedy algorithm to solve the joint content caching and recommendation problem with unknown user threshold. By the $\varepsilon$-greedy algorithm, in each time slot $t$, the BS either applies Algorithm 1 (with probability $1 - \varepsilon$) based on the current estimate $\hat {\boldsymbol{\theta}}^{(t)}$ to obtain the caching and recommendation policy, or recommends contents in the set $\mathcal{R}_u^{(t)} \triangleq \{f|p_{uf} < \hat\theta_{u}^{(t)} \}$ randomly for each user  (with probability $\varepsilon$). The details are provided in Algorithm \ref{alg:2}.

\begin{algorithm}
	\caption{\small $\varepsilon$-greedy algorithm}
	\label{alg:2}
	\renewcommand{\algorithmicrequire}{\textbf{Input:}}
	\renewcommand{\algorithmicensure}{\textbf{Output:}}
	\begin{algorithmic}[1] \small
		\Require $\{p_{uf}\}$, $\{v_u\}$, $\varepsilon$
		\State Initialize $\hat{\boldsymbol{\theta}}^{(1)}= \mathbf{1}$
		\For{time slot $t=1,2,3,\cdots$}
		\parState{Generate a uniformly distributed random variable ${\sf rand}\in[0,1]$.}
		\If{${\sf rand}  > \varepsilon$} \Comment{Exploitation Step}
		\parState{Obtain $\mathbf{M}^{(t)}$ and $\mathbf{c}^{(t)}$ by Algorithm 1 based on the estimated threshold $\hat{\boldsymbol{\theta}}^{(t)}$.}
		\Else \Comment{Exploration Step}
		\parState{Set $\mathbf{M}^{(t)}$ by recommending contents in $\mathcal{R}_{u}^{(t)}$ randomly and set $\mathbf{c}^{(t)} = {\sf c}(\mathbf{M}^{(t)},\hat{\boldsymbol{\theta}}^{(t)})$.}
		\EndIf
		\State Observe the user demands in time slot $t$.
		\For{each user-content request tuple $(u, f)$ during $t$}
		\If{$I(u,f) = 1$ and $\hat\theta_u^{(t)} > p_{uf}  $}
		\State Update $\hat\theta_u^{(t + 1)} \leftarrow p_{uf}$
		\EndIf
		\EndFor
		\EndFor
	\end{algorithmic}
\end{algorithm}

The convergence of the $\varepsilon$-greedy algorithm is shown in the following proposition.
\begin{proposition}
	The average number of time slots needed for $\hat\theta_u^{(t)}$ to converge to $\tilde \theta_u$ is upper bounded by
	\begin{equation}
	T < \bar T \triangleq \tfrac{N_f^2  }{\varepsilon}\Big(  N_m - N_m\big(1 -\tfrac{ \rho_u}{N_m}\big)^{N_q(u)}\Big)^{-1}
	\end{equation}
	where $N_q(u)$ is the number of requests of the $u$th user in each time slot, $\rho_u = \min\limits_{f\in\mathcal{F}_u}\{ p_{uf} \}$ and $\mathcal{F}_u = \{f|p_{uf} > 0\}$.
\end{proposition}
\begin{IEEEproof}
	See Appendix B.
\end{IEEEproof}

Since $\bar T$ is a finite value, we have $\lim_{t\to\infty}\hat{\boldsymbol{\theta}}^{(t)} = \tilde {\boldsymbol{\theta}}$. Then, based on Remark 1, the policy in the exploitation step of Algorithm 2 is the same as the oracle policy for $t\to\infty$. Considering that the exploitation probability is $1-\varepsilon$, Algorithm 2 achieves the successful offloading probability achieved by the oracle policy with probability $1 - \varepsilon$. Further considering that the successful offloading probability achieved by the exploration step of Algorithm 2 is at least zero, Algorithm 2 can achieve at least $1-\varepsilon$ of the performance achieved by the oracle policy on average when $t \to \infty$.
%As shown in the simulation, the convergence can be fast with proper chosen $\varepsilon$.

\section{Simulation Results}
In this section, we compare the performance of the proposed caching and recommendation policies with prior works, and analyze the impact of various factors by simulation.

The following baselines are considered for comparison.
\begin{enumerate}
	\item \emph{``Rec UP -- Cache Pop -- Rec Adj"}: This method first assumes that each user is recommended with the top-$N_m$ preferred contents according to individual inherent user preference, and then lets each BS cache the top-$N_c$ popular contents after recommendation and  adjusts recommendation list based on the cached contents. This is the policy proposed in \cite{nudge} when the contents are set with equal size.
	\item \emph{``Rec Pop -- Cache Pop"}: Each BS recommends the same $N_m$ contents to every user that maximize cache-hit probability. This can be served as the  performance upper bound of the policy proposed in \cite{guo}, which ignores the heterogeneity of inherent user preference and is not a personalized recommendation policy.
	\item \emph{``Cache Opt -- Rec UP"}: Each BS caches contents according to \eqref{eqn:c} but based on inherent user preference (i.e., replacing $q_{uf}$ with $p_{uf}$ in \eqref{eqn:c}) and recommends top-$N_m$ preferred contents to each user according to inherent user preference. This reflects existing strategy where recommendation and caching decisions are made independently.
	\item \emph{``Cache Opt -- No Rec"}: This is an existing optimal caching policy without recommendation \cite{liu2017caching}.

\end{enumerate}

The pathloss exponent and number of antennas are set as $\alpha = 3.76$ and $N_t = 2$, which are typical for a pico BS. The SIR requirement is set as $\gamma_0=  -8$ dB, which corresponds to 2 Mbps rate requirement with 10 MHz transmission bandwidth.
The inherent preference $p_{uf}$ and activity level $v_u$ are obtained from the logs of Million Songs dataset \cite{MSD}. To reduce the simulation time, we choose the top-50 active users and top-100 most-listened songs by these users from the dataset, i.e., $N_u=50$, $N_f=100$. Specifically, the inherent user preference is estimated as the ratio of the times of a song listened by a user to the total listening times of the user, and the activity level is estimated as the ratio of the total listening times of the user to the total listening times of all the users. The recommendation list size is $N_m = 5$ and the user thresholds are set as uniformly distributed random variables, $\theta_u \sim \big[0, \theta_{\max}]$. In Algorithm 1, we set $N_a = |\{f| p_{uf} \geq \theta_u \}| + N_m$ for the $u$th user. Unless otherwise specified, $\theta_{\max} = 4/N_f$ (by this setting, about 10\% of the  contents satisfy $p_{uf} \geq \theta_u$ per user on average).

We first assume that ${\boldsymbol{\theta}}$ is known \emph{a priori} for a fair comparison, since prior works assume known user demands after recommendation.

\begin{figure}[!htb]
	\vspace{-1mm}
	\centering
	\includegraphics[width=0.35\textwidth]{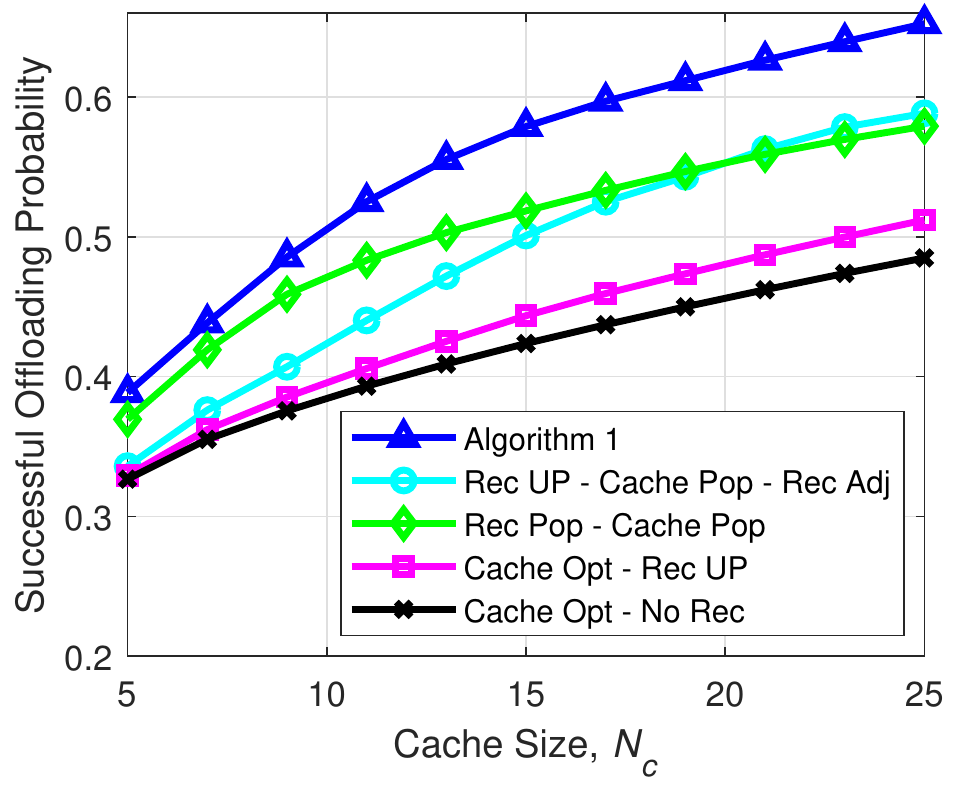}
	\vspace{-4mm}
	\caption{Successful offloading probability versus cache-size.} \label{fig:CHR_vs_Nc}
	\vspace{-1mm}
\end{figure}

In Fig. \ref{fig:CHR_vs_Nc}, we compare the performance of the proposed algorithm with the baseline policies. It is shown that Algorithm 1 outperforms all baseline policies.
\footnote{In fact, for small $N_u$ and $N_f$ where exhaustive searching is affordable, simulation results show that Algorithm 1 performs very close to the global optimal solution of $\mathsf{P}_1$ found by exhaustive searching. Due to space limitation, we do not provide the result for conciseness.} 
Compared with the policy in \cite{nudge}, Algorithm 1 increases $20\%$ of the successful offloading probability. Compared with the policy without recommendation in \cite{liu2017caching}, the gain is $40\%$.

\begin{figure}[!htb]
	\centering
	\includegraphics[width=0.35\textwidth]{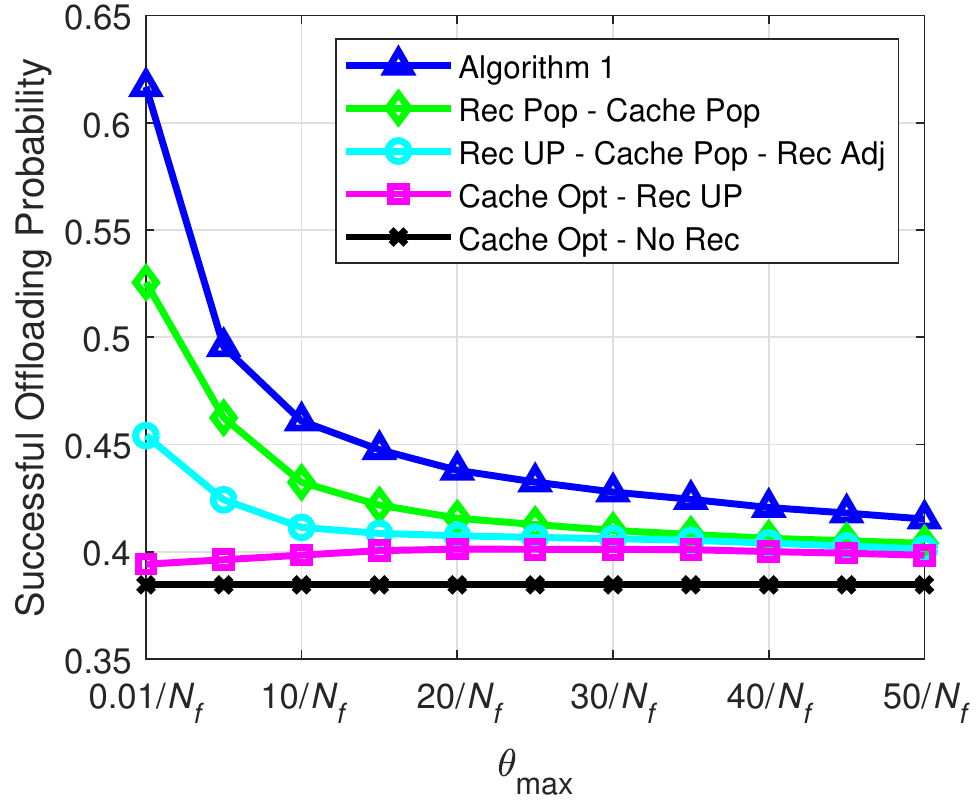}
	\vspace{-3mm}
	\caption{Successful offloading probability versus user threshold, $N_c = 10$.} \label{fig:CHR_vs_theta}
    \vspace{-1mm}
\end{figure}

In Fig. \ref{fig:CHR_vs_theta}, we show the impact of the user threshold. By setting $\theta_{\max}$ from $0.01/N_f$ to $50/N_f$, the percentage of contents satisfying $p_{uf} \geq \theta_u$ ranges from $15\%$ to $2\%$ per user on average. Except ``Cache Opt - Rec UP", the performance of the policies with recommendation decreases with the increase of threshold. This is because user demands are less affected by recommendation when the thresholds are high. ``Cache Opt - Rec UP" only slightly outperforms the caching policy without recommendation. This is because its caching and recommendation policies operate separately. Recommending the most preferred contents to each user may make the user preferences more heterogeneous, and hence the content popularity becomes less skewed.

\begin{figure}[!htb]
	\centering
	\includegraphics[width=0.35\textwidth]{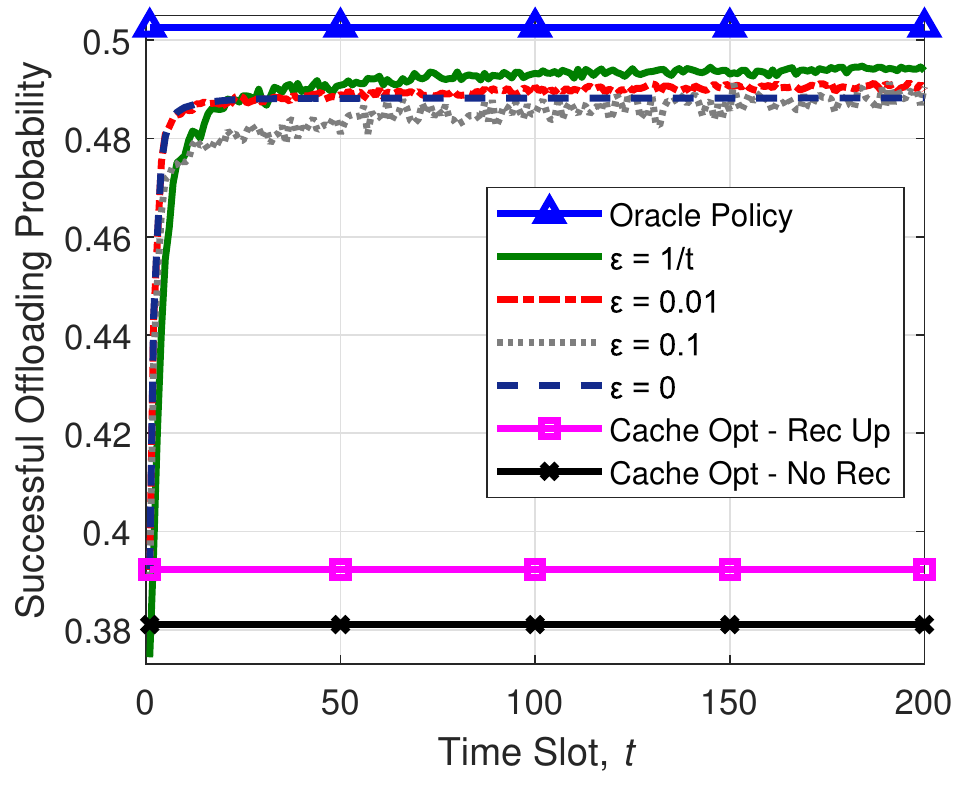}
	\vspace{-2mm}
	\caption{Convergence performance with learned user threshold, $N_c = 10$. In each time slot, $200$ user requests arrive randomly. } \label{fig:learning}
	\vspace{-2mm}
\end{figure}

In Fig. \ref{fig:learning}, we compare the performance of the proposed $\varepsilon$-greedy algorithm with existing methods when the user threshold is unknown. Note that baselines 1) and 2) are  not applicable in this case since they require the knowledge of user demands after recommendation. To show the tradeoff between exploration and exploitation, we compare different values of $\varepsilon$. When $\varepsilon = 0$, the algorithm always exploits but never explores and its performance reaches plateaus quickly. As $\varepsilon$ increases, the probability for exploration increases. When $\varepsilon = 0.01$, the algorithm can be improved continuously over time with exploration (though relatively slow). The algorithm with $\varepsilon = 0.1$ learns the user threshold more quickly due to higher exploration probability, but only exploits in 90\% of the time slots and hence the resulting performance is inferior to the algorithm with $\varepsilon = 0.01$.
When exploring more at the beginning and then reducing $\varepsilon$ properly over time, e.g., by setting $\varepsilon = 1/t$, the algorithm can achieve a good balance between learning faster and performing better.

\section{Conclusion}
In this paper, we jointly optimized content caching and recommendation at base stations with known and unknown user preference after personalized recommendation. We provided a model to character the user preference after recommendation by introducing a psychological threshold reflecting user personality. We proposed a hierarchical iterative algorithm to solve the optimization problem with known threshold, and proposed an $\varepsilon$-greedy algorithm to find the solution by learning the threshold. Simulation results showed that the algorithms can improve successful offloading probability significantly compared with existing solutions. The $\varepsilon$-greedy algorithm can converge quickly to achieve more than $1-\varepsilon$ of the performance obtained by the hierarchical iterative algorithm with perfect user threshold.

\appendices
\section*{Appendix A: The derivation of $\mathbb{P}(\gamma_{f} > \gamma_0)$}
Based on the law of total probability, we can obtain
\begin{equation}
\mathbb{P}(\gamma_{f} > \gamma_0) = \textstyle\int_{0}^{\infty}\mathbb{P}(\gamma_{f} > \gamma_0 ~|~ r) f_r(r) dr  \label{eqn:P}
\end{equation}
where $f_r(r) = e^{- c_f \lambda \pi r^2} 2\pi c_f\lambda r$ is the probability density function of the distance between the user
and its serving BS when requesting the $f$th content, and $\mathbb{P}(\gamma_{f} > \gamma_0 ~|~ r)$ is the conditional success probability. From \eqref{SIR}, we have
\begin{align}
\mathbb{P}(\gamma_{f} > \gamma_0 ~|~ r) &  = \mathbb{E}_{I_f, I'_{f}}[\mathbb{P}[ h > N_t r^{\alpha} (I_f +I'_{f})\gamma_0~|~r,I_f, I'_{f} ] ] \nonumber \\
& \overset{(b)}{=} \mathbb{E}_{I_f} \big[e^{-N_t r^{\alpha}\gamma_0 I_f}\big] \mathbb{E}_{I'_{f}} \big[e^{-N_t r^{\alpha}\gamma_0I'_{f}}\big] \label{eqn:P1}
\end{align}
where the last step is from $h\sim\exp(1)$ and the fact that $I_f$ and $I'_f$ are independent. Then, we can derive
\begin{align}
& \mathbb{E}_{I_f} [e^{-N_t r^{\alpha}\gamma_0 I_f}]  = \mathbb{E}_{\Phi_{f}, \{g_b\}} [e^{-N_t r^{\alpha}\gamma_0 \sum_{b\in\Phi_f} g_b r_b^{-\alpha}}] \nonumber \\
& \overset{(a)}{=} \mathbb{E}_{\Phi_{f}} \Big[\textstyle\prod_{b\in\Phi_{f}\backslash b_0} \big(1 + \gamma_0 r^{\alpha}r_b^{-\alpha}\big)^{-N_t}\Big] \nonumber \\
& \overset{(b)}{=} e^{-2\pi c_f \lambda \int_{r}^{\infty} (1 - (1 + \gamma_0 r^{\alpha}r_b^{-\alpha} )^{-N_t} ) r_b {\rm d} r_b }  \nonumber \\
& = e^{-\pi c_f \lambda r^2 ({}_{2}F_1[-\frac{2}{\alpha}, N_t; 1-\frac{2}{\alpha}, -\gamma_0 ] - 1 ) } \label{eqn:L1}
\end{align}
where step $(a)$ follows from $g_b \sim \mathbb{G}(N_t , 1/N_t )$, step $(b)$
is from using the probability generating function of the PPP.

Similar to the derivation of \eqref{eqn:L1}, but considering that the BSs not caching the $f$th content can be arbitrarily close to the user, we can obtain
\begin{align}
& \mathbb{E}_{I'_f} [e^{-N_t r^{\alpha}\gamma_0 I'_f}] = e^{-2\pi (1 - c_f) \lambda \int_{0}^{\infty} (1 - (1 + \gamma_0 r^{\alpha} r_b^{-\alpha} )^{-N_t}  )  r_b {\rm d}r_b }\nonumber \\
&  =  e^{-\pi (1 - c_f)\lambda r^2 \Gamma (1-\frac{2}{\alpha}) \Gamma(N_t + \frac{2}{\alpha}) \Gamma(N_t)^{-1}\gamma_0^{\frac{2}{\alpha}} }  \label{eqn:L2}
\end{align}
Finally, by substituting \eqref{eqn:L1} and \eqref{eqn:L2} into \eqref{eqn:P1} and then into \eqref{eqn:P}, we can obtain the expression of $\mathbb{P}(\gamma_{f} > \gamma_0)$.
%\section*{Appendix B: Proof of Proposition 1}
%Consider a special case of $\mathsf{P}_1$ when $\theta_u = 0, \forall u$ and with fixed caching policy $\mathbf{c}^{(t)}$. Then, we have $a_{uf} = 1, \forall u,f$, and by substituting \eqref{eqn:quf} into \eqref{eqn:s},  $\mathsf{P}_1$ degenerates into a recommendation optimization problem
%\begin{subequations}
%	\begin{align}
%	\mathsf{P}_2:~ \quad \max_{\mathbf{M}^{(t)}} ~& \sum_{u=1}^{N_u}v_u  \frac{\sum_{f=1}^{N_f} m_{uf}^{(t)}  p_{uf}\mathbb{P}(\gamma_f \geq \gamma_0)}{\sum_{f' = 1}^{N_f}  m_{uf'}^{(t)}p_{uf'}}  \\
%	s.t. ~& \sum_{f = 1}^{N_f} m_{uf}^{(t)} = N_m, \forall u
%	\end{align}
%\end{subequations}
%which is  a \emph{product assortment problem}  with mixtures of logits that is NP-hard~\cite{assortment}. Considering that $\mathsf{P}_2$ is a simplified special case of $\mathsf{P}_1$, $\mathsf{P}_1$ is also NP-hard.

\section*{Appendix B: Proof of Proposition 1}
We consider the worst case scenario when $\theta_u \leq \min_{f}\{p_{uf}\}$ and $\hat \theta_u^{(t)}$ is only updated in the exploration step. In this case, the number of updates required to obtain $\hat\theta_u^{(t)} = \tilde\theta_u$ is at most $N_f$. Denote $n_u(t) = 1$ if $\hat\theta_u^{(t)}$ is updated in the $t$th time slot, otherwise $n_u(t) = 0$. Define $T(\delta_u)$ as the number of time slots needed to update $\hat\theta_u^{(t)}$ for $N_f$ times with update probability $\delta_u$ in each time slot. Then, according to Wald's Equation in martingale theory \cite{martingale}, we can obtain $
N_f = \mathbb{E}\big[\sum_{t=1}^{T(\delta_u)} n_u(t) \big] = \mathbb{E}[T(\delta_u)] \mathbb{E}[n_u(t)] = \mathbb{E}[T(\delta_u)] \delta_{u}$, from which we have
\begin{equation}
\mathbb{E}[T(\delta_u)] = {N_f}/{\delta_u} \label{eqn:T}
\end{equation}

According to Remark 2, to update $\hat \theta_u^{(t)}$ in the $t$th time slot, the recommendation list should contain at least one content, say the $f$th content, that satisfying $p_{uf} < \hat \theta_u^{(t)}$, and the user requests the $f$th content from the recommendation list. If $\hat\theta_u^{(t)}$ is only updated in exploration step, $\delta_u$ is lower bounded by
\begin{equation}
\delta_u \geq \varepsilon \tbinom{N_f -1}{N_m-1}\tbinom{N_f}{N_m}^{-1} \Big(1 - \big(1 - \tfrac{ \rho_u}{N_m}\big)^{N_q(u)}\Big) \label{eqn:lower}
\end{equation}
where $\binom{N_f -1}{N_m-1}\binom{N_f}{N_m}^{-1}$ is the lower bound of the probability that the recommendation list contains only one content $f$ satisfying $p_{uf} < \hat \theta_u^{(t)}$, $\frac{\rho_u}{N_m}$ is the lower bound of the probability that the $u$th user requests the $f$th content from the recommendation list according to the user demands model, and hence $1 - \big(1 - \frac{ \rho_u}{N_m}\big){}^{N_q(u)}$ is the lower bound of the probability that the $u$th user requests the $f$th content from the recommendation list in the $t$th time slot.
Then, by substituting \eqref{eqn:lower} into \eqref{eqn:T} and with some manipulations, Proposition 1 can be proved.
\bibliographystyle{IEEEtran}

\bibliography{dongbib}

\end{document}